# Coherent Evolution of Bouncing Bose-Einstein Condensates

K. Bongs,[1] S. Burger,[1] G. Birkl,[1] K. Sengstock,[1] W. Ertmer,[1] K. Rzążewski,[2,3] A. Sanpera,[3] and M. Lewenstein[3]

[1]*Institut für Quantenoptik, Universität Hannover, 30167 Hannover, Germany*
[2]*Centrum Fizyki Teoretycznej, Polska Akademia Nauk, 02668 Warsaw, Poland*
[3]*Institut für Theoretische Physik, Universität Hannover, 30167 Hannover, Germany*
(Received 4 March 1999)

We investigate the evolution of Bose-Einstein condensates falling under gravity and bouncing off a mirror formed by a far-detuned sheet of light. After reflection, the atomic density profile develops splitting and interference structures which depend on the drop height, on the strength of the light sheet, as well as on the initial mean field energy and size of the condensate. We compare experimental results with simulations of the Gross-Pitaevski equation. A comparison with the behavior of bouncing thermal clouds allows us to identify quantum features specific for condensates.

Since the first experimental realization of Bose-Einstein condensation (BEC) in weakly interacting atomic systems [1–3], many fundamental experiments with BEC's have been performed [4]. One of the most interesting future prospects for Bose-Einstein condensates is their application as a source of coherent matter waves [5], e.g., in atom optics and atom interferometry. This promises a significant advance similar to the introduction of lasers in light optics. The application of coherent matter waves in phase sensitive experiments, like interferometers, necessitates the understanding of their evolution when being manipulated by atom optical elements like mirrors and beam splitters. The dynamics of coherent matter waves during and after the interaction with these elements is in comparison to single-atom optics much more complex and can be termed "nonlinear atom optics." As one of the key elements, atom mirrors [6] deserve a detailed investigation.

In this Letter, we report on the bouncing of atomic BEC's off a mirror formed by a repulsive dipole potential. Condensates released from a magnetic trap fall under the influence of gravity and interact with a blue-detuned far-off-resonant sheet of light. A spatial splitting of the reflected ensemble and the appearance of self-interference substructures are observed. Surprisingly, for certain conditions thermal clouds also exhibit splitting into two parts after bouncing. We compare our experimental results with solutions of the Gross-Pitaevski equation for the condensate and with the classical Liouville equation for the thermal cloud. The comparison shows that, although condensate splitting has its classical counterpart, it exhibits genuine quantum features.

In our experimental setup, condensates typically containing $10^5$ $^{87}$Rb atoms in the ($F = 2$, $m_F = +2$)-state are produced every 20 s. It has been checked that less than 10% of the atoms of the cloud are not in the condensate fraction. This corresponds to a temperature range of $\simeq 200$ nK. The fundamental frequencies of our magnetic trap (a "cloverleaf" trap [7]) are $\omega_{\parallel} = 2\pi \times 17$ Hz and $\omega_{\perp} = 2\pi \times 340$ Hz along the axial and radial directions, respectively. Therefore, the condensates are pencil shaped with the long axis oriented horizontally. Our trap can be switched off within 200 $\mu$s, and—after a variable delay time—the density distribution of the atom sample can be detected using absorption imaging, dark field imaging, or phase-contrast imaging. The 2D-image plane contains the weak trap axis as well as the strong trap axis along the direction of gravity. Further details of our experimental setup will be presented elsewhere.

A repulsive optical dipole potential is created by a far off-resonant Gaussian laser beam with a wavelength of $\lambda = 532$ nm and a power of up to 3 W, focused to a waist of about 10 $\mu$m. The beam is spatially modulated with an acousto-optic deflector in the horizontal plane. The modulation period of typically 10 $\mu$s is much shorter than the time the atoms spend inside the dipole potential. For the atoms this results in a time-averaged static dipole potential. In contrast to magnetic mirrors or evanescent wave mirrors, changing the modulation waveform gives the flexibility to externally define the intensity profile. For the experiments presented here, we use a flat light sheet, oriented nearly perpendicular to the direction of gravity. The time-averaged beam profile has a spatial extent of $\simeq 200$ $\mu$m in the horizontal direction. The interaction of the atoms with the light field is dominated by the dipole transitions at a wavelength of 780 nm and 795 nm, leading to a repulsive potential barrier with a detuning large enough to make spontaneous emission negligible. Atoms, dropped from a height of up to 300 $\mu$m can thus totally be reflected.

In several series of measurements we have studied in detail the evolution of BEC's bouncing off the atom mirror. We recorded atomic clouds being reflected up to 3 times off the mirror before laterally moving out of the field of view due to a slight slope in the orientation of the light sheet. The behavior of the reflected atomic samples could be modified by changing the drop height, as well as the power and waist of the light sheet. This allowed for the observation of two regimes in the wave packet dynamics: the dispersive reflection off weak dipole potentials ("soft" mirror) and the nearly nondispersive reflection off strong dipole potentials ("hard" mirror).

Figure 1(a) shows a time-of-flight series of Bose-Einstein condensates bouncing off a soft mirror. Each frame is recorded with a different condensate, created under identical experimental conditions. The light sheet is positioned 270 $\mu$m below the magnetic trap and shows up as the sharp lower edge in the fourth frame. The high kinetic energy accumulated before hitting the atom mirror causes the atomic cloud to penetrate deep into the dipole potential before being reflected. The corresponding classical turning point is situated near the maximum of the Gaussian intensity profile, in a region with a weak gradient of the repulsive potential. Further increasing the drop height or reducing the laser power results in a partial transmission through the mirror.

As the condensate reapproaches the initial altitude, it splits and develops substructures [frames 6 to 11 in Figs. 1(a) and 2(a)]. We checked by independent measurements that the interferences are not due to our detection optics or detection method and do not occur for temperatures above the critical temperature for BEC, $T_c$. As we show below, the interference structures prove the persistence of matter-wave coherence for BEC's being reflected off the dipole potential atom mirror.

In another set of measurements, a nearly nondispersive mirror for Bose-Einstein condensates was created by placing an intense light sheet closer to the magnetic trap (155 $\mu$m drop height). This resembles reflection off an infinite potential step. The temporal evolution of the condensate after reflection off such a hard mirror is presented in Fig. 3. Close to the upper turning point, the atom cloud is refocused to a narrow distribution along the direction of gravity [Fig. 3(c)], and develops into a double-peak structure shortly after the upper turning point [Fig. 3(d)]. No interference structures such as those presented in Fig. 1(a) are observed here.

In order to compare the bouncing of condensates to single-atom optical effects, we have performed bouncing experiments with atom clouds cooled to a temperature just above the critical temperature, $T_c$. Surprisingly, thermal clouds of ultracold atoms also reveal splitting after reflection off a hard atom mirror. Typically, the splitting is larger than for BEC's [see Fig. 1(b) and 2].

This poses two questions: First, what is the physical mechanism underlying the splitting of thermal clouds? Second, to which extent do the structures observed in condensate bouncing reveal quantum behavior?

In order to answer the first question we have studied classical dynamics of a cloud of noninteracting atoms bouncing off a hard mirror. The Liouville equation for the flow of density in phase space can be solved analytically or simulated by using Monte Carlo phase space averaging over the initial state. Indeed, classical dynamics leads to splitting of the cloud right after passing through the upper turning point.

Initially, velocity and space distributions are completely decorrelated. During the free fall, velocity-space correlations arise—particles with velocities downwards initially are located lower than those with velocities upwards initially. The wave packet undergoes, however, nothing but the spreading. First after the upper turning point, both groups tend to refocus, and separate. One can simply estimate that their separation at this point will be $\sim(\Delta v)^2/g$, where $\Delta v$ is the initial velocity spread and $g$ is the gravitational acceleration. For a splitting to occur this separation has to be larger than the widths of the two groups of particles which at the upper turning point are close to the initial width, $\Delta z$, of the atom cloud, i.e., $(\Delta v)^2/g\Delta z > 1$. Our experimental results for thermal clouds agree well with this simple theory and with numerical simulations.

To answer the second question, we have performed numerical simulations to mimic the experimental behavior of the condensate. A simple estimate shows that the initial condensate wave function, $\Psi(x, y, z)$, which can be reproduced using the Thomas-Fermi approximation [8], leads to a density distribution of width $\Delta z \sim$ few $\mu$m. The velocity spread is thus of the order of $\Delta v \simeq \hbar/M\Delta z \simeq 10^{-1}$ $\mu$m/ms, with atom mass $M$. A classical cloud with such parameters does not show any splitting or specific structure. Similarly, the wave packet of a single atom evolving according to the linear Schrödinger equation does

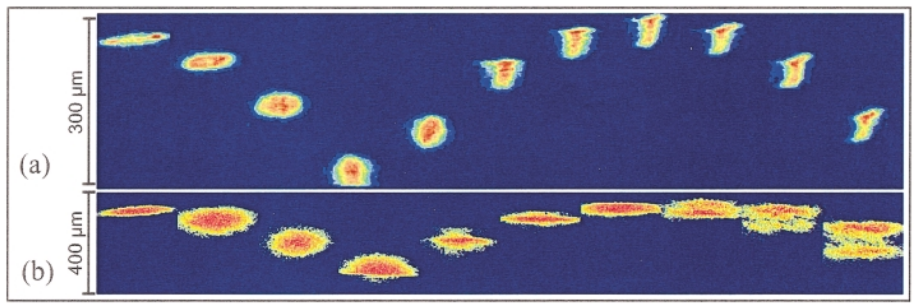

FIG. 1 (color). (a) Series of dark field images for condensates bouncing off a light sheet 270 $\mu$m below the magnetic trap. Each image was taken with a new condensate and with an additional time delay of 2 ms. The density of the condensate during the first few ms of expansion causes a phase shift in the detection light of more than $2\pi$, which explains the stripes in the middle of the first two images. (b) A thermal cloud bouncing off a light sheet situated 230 $\mu$m below the magnetic trap splits into two parts.

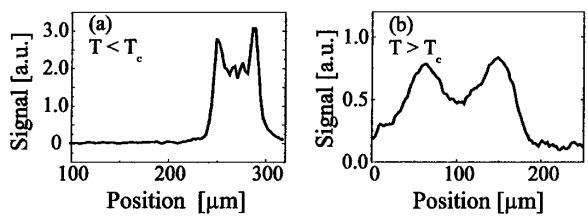

FIG. 2. Cross sections of images of a bouncing BEC (a) and a bouncing thermal cloud (b). Graph (a) corresponds to frame 9 of Fig. 1(a). Graph (b) is a cross section of the last frame in Fig. 1(b).

not split. In the condensate, however, the potential energy of the atom-atom interaction is transferred into kinetic energy within few ms of ballistic expansion, allowing thus for the splitting.

In principle, the simulations require solutions of the 3D Gross-Pitaevski equation, which is a serious numerical task for the considered regime of length and time scales. We believe, however, that one can reduce the problem to the vertical direction, because the dynamics along the horizontal directions consists in free expansion essentially. With decreasing condensate density, the evolution becomes linear (i.e., well described by the linear Schrödinger equation) and remains such. In order to incorporate the effects of ballistic expansion of the condensate into the 1D simulations we adiabatically turn off the nonlinearity within a few ms of the time evolution. Thus we solve the Gross-Pitaevski equation in 1D,

$$i\hbar \frac{\partial}{\partial t} \psi(z) = \left[ -\frac{\hbar^2}{2M} \frac{d^2}{dz^2} + V(z) + Nu(t)|\psi(z)|^2 \right] \psi(z),$$ (1)

where $V(z)$ describes the gravitational potential and the potential of the light sheet that acts as a mirror. The latter is assumed to form a Gaussian barrier of a height $V_b \sim 280$ $\mu$m $\times Mg$, and a width of $\simeq 5-10$ $\mu$m. The form of the potential plays a role in the dynamics when the initial gravitational energy becomes comparable to the light sheet barrier $V_b$.

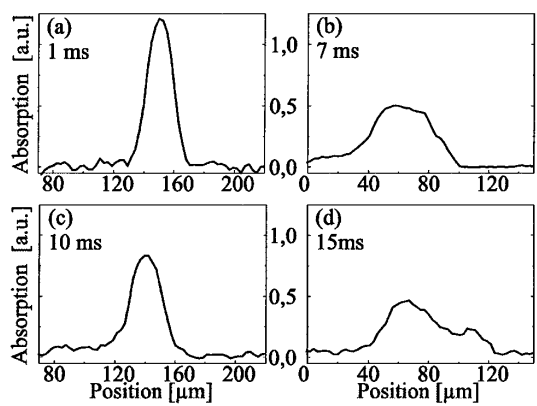

FIG. 3. Cross sections of time-of-flight absorption images for condensates bouncing off a light sheet 155 $\mu$m below the magnetic trap. In (c) the reflected atom cloud is near the upper turning point and exhibits refocusing along the vertical dimension. After 15 ms (d) the condensate splits into two parts.

In Eq. (1), $Nu(t)|\psi(z)|^2$ represents the mean field energy of atom-atom interactions, and $N$ is the total number of particles in the condensate. The function $u(t)$ mimics the expansion of the condensate along the horizontal directions including the dramatic reduction of the interparticle energy. We model this by writing $u(t) = u(0)\exp(-t/\tau)$, with $\tau = 4$ ms. The results do not depend significantly on the shape of the dynamical switching off of $u(t)$. The value of $u(0)$ is fixed by requiring that the total potential energy of atom-atom interactions in the 1D simulation at $t = 0$ equals the total potential energy of atom-atom interactions calculated from the 3D Gross–Pitaevski equation, i.e., $u(0) \int dz \times |\psi(z)|^4 = u_0 \int dx\,dy\,dz\,|\Psi(x,y,z)|^4$, where $u_0 = 4\pi\hbar^2 a_s/M$ and $a_s \simeq 5.8$ nm is the scattering length for rubidium atoms. The nonlinear energy in 3D can be estimated using the Thomas-Fermi approximation [8].

As initial wave function of the condensate we consider a Gaussian wave packet, $|\psi(z)|^2 = \int dx\,dy \times |\Psi(x,y,z)|^2 \propto \exp -(z^2/2R^2)$, where $R$ is the radius of the condensate in the 3D Thomas-Fermi approximation. The value of the nonlinear coupling reads $u(0) = (4\pi\hbar^2 a_s \lambda/MR^2)(15/7\sqrt{\pi})$, with $\lambda = \omega_{\parallel}/\omega_{\perp}$. Using the experimental values of $\omega_{\parallel}$ and $\omega_{\perp}$, we obtain a coupling value of $u(0)/\hbar \simeq 9 \times 10^{-4}$ $\mu$m/ms. As an important result, the splitting of the condensate is very sensitive to the value of the nonlinearity—therefore, its appearance allows one to estimate the atom number $N$. For the value of the nonlinear coupling given above, splitting is observed only for atom numbers $N \geq 5 \times 10^4$. This is in good agreement with the experimental results, as the splitting vanished for small condensates produced by lowering the final RF of the evaporation cycle.

The results of the simulations are presented in Fig. 4 for a hard mirror and in Fig. 5(b) for a dispersive soft mirror. In both cases, the time evolution resembles to some extent the classical results, but the physics is quite different. The condensate wave packet spreads quite rapidly during the first few ms, and this spreading is entirely due to nonlinear interactions. At the upper turning point, the wave packet squeezes back to a few microns width, and right afterwards, it undergoes splitting [see Fig. 4(d)]. This splitting is significantly smaller than in the classical case. In the latter case, a large velocity spread of a thermal cloud above $T_c$ causes the splitted parts to move apart very fast, while in the former case, the splitted parts do not move apart so quickly.

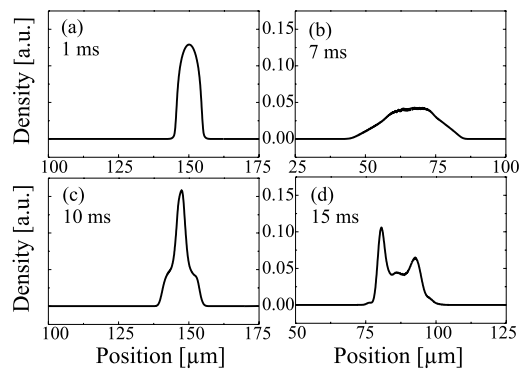

FIG. 4. Snapshots of the numerical simulation corresponding to Fig. 3, number of atoms $N = 5 \times 10^4$.

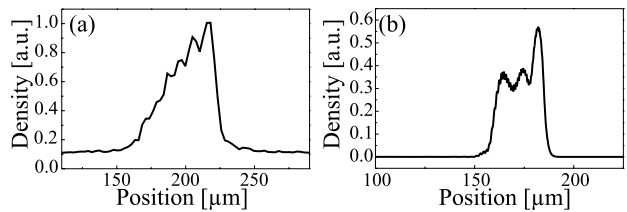

FIG. 5. Experimentally observed self-interference structures in the condensate (a) and numerical results (b) for the same parameters as in the experiment ("soft" mirror 270 $\mu$m below the magnetic trap, $N = 8 \times 10^4$, time evolution of 11 ms).

In agreement with the experimental results, the numerical results also show additional structures before and after the upper turning point when a soft mirror is placed 270 $\mu$m below the trap (Fig. 5). Evidently, when the initial gravitational energy is comparable to the mirror height, the softness of the mirror causes velocity dependent dispersion for the reflected matter waves, leading to interferences and density modulation.

We have calculated the condensate phase-space Wigner functions, for soft and hard mirrors. For both cases, the Wigner functions oscillate between positive and negative values in the central region of the space (i.e., close to mean position and momentum of the wave packet). The interference structure can be explained by the fact that the splitting parts of the wave packet overlap in the central region, where the particles have positions and velocities very close to the mean ($\Delta v \sim 0.5$ mm/s, which corresponds to a fringe separation of $h/M\Delta v \sim 10~\mu$m, as observed). Each of the two parts has a different spatial phase dependence, and interference is observed. This effect is enhanced for soft mirrors. Our analysis indicates, indeed, that in the quantum regime splitting cannot be regarded as a purely classical effect.

Our numerical results agree well with the experimental observations and clearly explain the appearance of the different splitting behavior for noncondensed samples and BEC's as well as the self-interference structure for bouncing off a "soft" mirror. In a direct comparison, the width of the experimental structures is broader for small structures than the calculated width, which is partly due to our limited spatial detection resolution ($\approx 10~\mu$m) and partly due to the thermal cloud surrounding the BEC, not considered in the calculations.

In conclusion, we have investigated the dynamics of Bose-Einstein condensates bouncing off an atom mirror. We have compared the behavior for different regimes in the ratio of kinetic energy to mirror strength as well as the difference between condensed and noncondensed atom samples. Both experimental results and numerical simulations based on the Gross-Pitaevski equation have been presented. For BEC's, splitting and interference structures in the atom density distribution are observed after reflection. In addition, a splitting has been found also for thermal atom samples with small initial spatial and large initial momentum spread.

We have presented the application of conservative optical potentials for the manipulation of coherent matter waves. The observation of splitting and of interference can be used to characterize and determine mirror properties such as roughness and steepness, and coherence properties of the condensate.

In addition to creating an atom mirror with reflectivity close to unity as shown here, we also have investigated partially reflecting mirrors and a phase shifter for coherent matter waves, where the optical potential delays the atoms but does not cause reflection. In future work, we will apply these results to develop atom interferometers for Bose-Einstein condensates. The observed coherent splitting itself may also be applied to realize an atom interferometer in a pulsed scheme, e.g., by the application of additional light fields acting as mirrors and phase shifters for the individual partial waves. We have shown that mirrors based on optical potentials can serve as a detection scheme for matter-wave coherence, i.e., the onset of BEC or the output properties of an atom laser. They may even be used to systematically characterize the coherence properties of these sources, or of coherent matter waves being manipulated by other techniques.

We thank P. Villain for fruitful discussions. K. R. thanks the Alexander von Humboldt-Foundation for its generous support. This work is supported by the SFB 407 of the *Deutsche Forschungsgemeinschaft*.